%% file: Prismatoids.tex
\documentclass{article}  
\usepackage{url}
\usepackage{graphicx}
\usepackage{amsfonts}
\usepackage{amssymb}
\usepackage{amsmath}
\usepackage{latexsym}
\usepackage{subfigure}

\input{mac}

\def\P{{\mathcal P}}

\begin{document}

\title{Unfolding Smooth Prismatoids}

\author{%
Nadia Benbernou%
   \thanks{Department of Mathematics 
      \protect\url{nbenbern@email.smith.edu}.
}
\and
Patricia Cahn%
   \thanks{Department of Mathematics 
      \protect\url{pcahn@email.smith.edu}.
}
\and
Joseph O'Rourke%
    \thanks{Department of Computer Science, Smith College, Northampton, MA
      01063, USA.
      \protect\url{orourke@cs.smith.edu}.
       Supported by NSF Distinguished Teaching Scholars award
       DUE-0123154.}
}
\date{}
\maketitle

\begin{abstract}
We define a notion for unfolding smooth, ruled surfaces,
and prove that every smooth prismatoid (the convex hull of two smooth curves
lying in parallel planes), has a nonoverlapping ``volcano unfolding.''
These unfoldings keep the base intact, unfold the sides outward, splayed
around the base, and attach the top to the tip of some side rib.
Our result answers a question for smooth prismatoids
whose analog for polyhedral prismatoids
remains unsolved.
\end{abstract}

\section{Introduction}
\seclab{Introduction}
It is a long-unsolved problem to
determine whether or not
every convex polyhedron can be cut along its
edges and unfolded flat into the plane to a single
nonoverlapping simple polygon (see, e.g.,~\cite{o-fucg-00}).
These unfoldings are known as \emph{edge unfoldings} because the surface
cuts are along edges;
the resulting polygon is called a \emph{net} for the polyhedron.
Only a few classes of polyhedra are known to have such unfoldings:
pyramids, prismoids, and domes~\cite{do-fucg-04}.
Even for the relatively simple class of prismatoids, nonoverlapping
edge unfoldings are not established.
(All these classes of polyhedra, except domes, will be defined below.)
In this paper, we generalize edge unfoldings to certain piecewise-smooth ruled surfaces,%
\footnote{
  A \emph{ruled surface} is one that can be swept out by a line moving in space.
  A patch of a ruled surface may therefore be viewed as composed of line segments.
}
and show that smooth prismatoids can always be unfolded without overlap.
Our hope is that the smooth case will inform the polyhedral case.

\paragraph{Pyramids.}
We start with pyramids and their smooth analogues, cones.
A \emph{pryamid} is a polyhedron that is the convex hull of a convex
\emph{base} polygon $B$ and a point $v$, the \emph{apex}, above the plane
containing the base.  The \emph{side faces} are all triangles.
It is trivial to unfold a pyramid without overlap:
cut all side edges and no base edge.
This produces what might be called a \emph{volcano} unfolding
(it blows out the side faces around the base).
Examples are shown in Fig.~\figref{reg.pyr.stars}(a,b) for regular
polygon bases.
\begin{figure}[htbp]
\centering
\includegraphics[width=\linewidth]{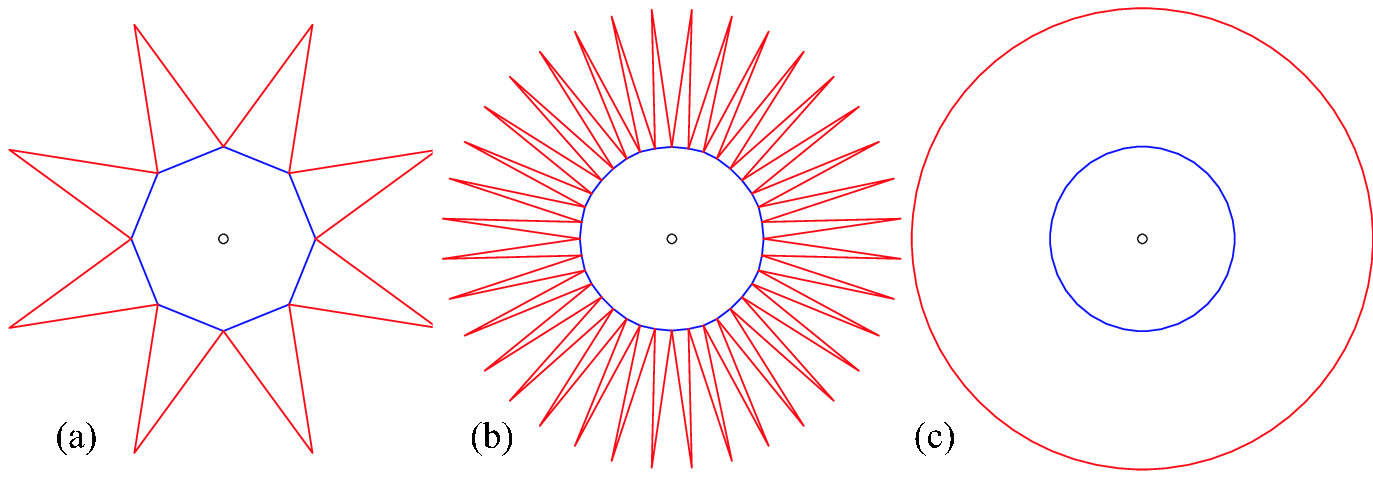}
\caption{Unfoldings of regular pyramids~(a-b) approaching the unfolding
of a cone~(c).}
\figlab{reg.pyr.stars}
\end{figure}

\paragraph{Cones.}
We generalize pyramids to \emph{cones}: shapes that
are the convex hull of a smooth convex curve base $B$ lying
in the $xy$-plane, and a point apex $v$ above the plane.
We define the volcano
unfolding of a cone to be the natural limiting shape as the number
of vertices of base polygonal approximations goes to infinity, and each side
triangle approaches a segment \emph{rib}.
This limiting process is
illustrated in Fig.~\figref{reg.pyr.stars}(c).
For any point $b \in \partial B$, the segment $vb$ is unfolded
across the tangent to $B$ at $b$.
Note that this net for a cone is no longer an unfolding that could be produced
by paper, because the area increases.  
(For a right circular cone of unit-radius base and unit height,
the surface area of the side of the cone is $2 \pi$, but the
area of the unfolding annulus is $\pi( 2^2 - 1^2 ) = 3\pi$.)
In a sense that can be made precise,
the density of the paper is thinned toward the tips of the spikes,
so that the integral of this density is the paper area unfolded.

\paragraph{Truncated Pyramids.}
The first extension of pyramids is to \emph{truncated pyramids}
those whose apexes are sliced off by a plane parallel to the base.
A cone truncation parallel to the base
produces the smooth analog;
see Fig.~\figref{truncatedcone}.
\begin{figure}[htbp]
\centering
\includegraphics[width=.5\linewidth]{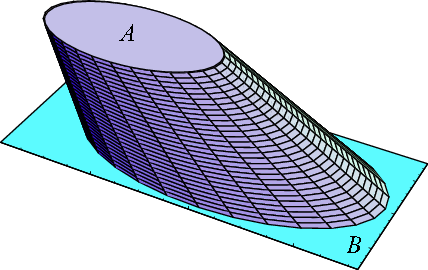}
\caption{Cone truncated by plane parallel to base.}
\figlab{truncatedcone}
\end{figure}
The goal now is to perform a volcano unfolding, with the addition
of attaching the top $A$ to the top of some side face for
a truncated polyhedron, or to the end of some side segment
for a truncated cone.
Fig.~\figref{mouseunfolding} shows the volcano unfolding
of a truncated cone with an irregular base, 
and two possible locations for placement of the top,
one that overlaps and one that does not.
Note that the boundary of the unfolding $U$ is not convex,
a point to which we will return.
\begin{figure}[htbp]
\centering
\includegraphics[width=0.6\linewidth]{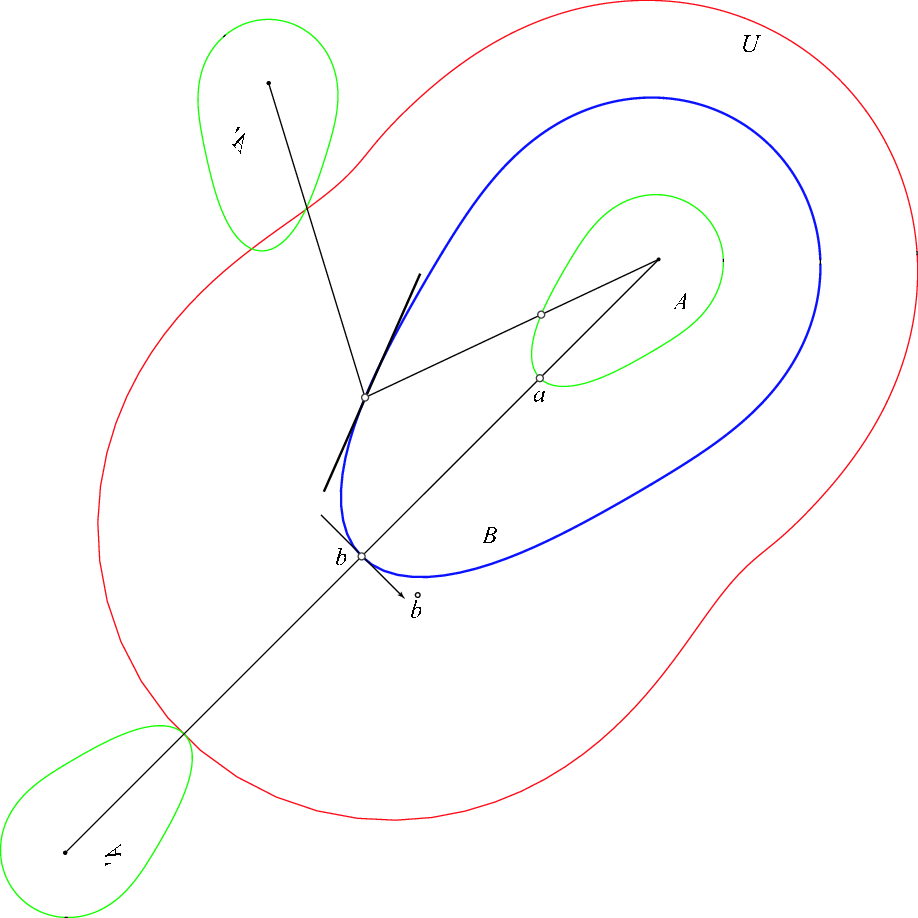}
\caption{A volcano unfolding of a truncated cone with two possible placements of top $A$
``flipped-out'' to $A'$.
$U$ is the boundary of the unfolding.}
\figlab{mouseunfolding}
\end{figure}

The next step in generalization is to polyhedra
known as \emph{prismoids}, of which a truncated pyramid is
a special case.  A prismoid can be defined 
as the convex
hull of two convex polygons $A$ and $B$ lying in parallel planes,
with $A$ angularly similar to $B$.
This last condition ensures that the side faces are trapezoids, each
with an edge of $A$ parallel to an edge of $B$.
An algorithm for edge-unfolding prismoids is available~\cite{do-fucg-04}.
It is a volcano unfolding, with the top $A$ attached to one carefully chosen
side face.  See Fig.~\figref{prismoid.unfolding}.
\begin{figure}[htbp]
\centering
\includegraphics[width=0.75\linewidth]{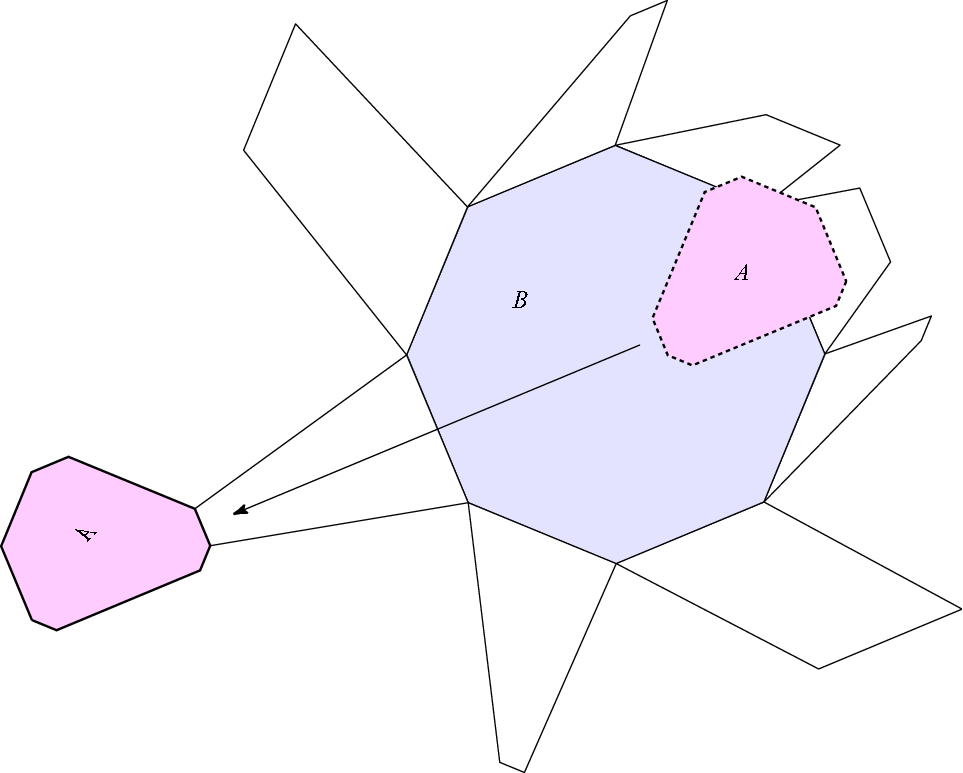}
\caption{Overhead view of a prismoid unfolding.  [From~\protect\cite{do-fucg-04}.]}
\figlab{prismoid.unfolding}
\end{figure}

\paragraph{Prismatoids}
The natural generalization of a prismoid is a \emph{prismatoid},
the convex hull of two convex polygons $A$ and $B$ lying in parallel planes,
with no particular relationship between $A$ and $B$.
As mentioned above, there is no algorithm for edge-unfolding prismatoids.
One complication for a volcano unfolding is that the side faces are generally
triangles, with base edges either on $B$ or on $A$.
Our concentration in this paper is on \emph{smooth prismatoids}, which we
define as the convex hull of two smooth convex curves $A$ above and $B$ below,
lying in parallel planes.
By \emph{smooth} we mean $C^2$: possessing continuous first and second derivatives.
A volcano unfolding of a smooth prismatoid unfolds every rib segment
$ab$ of the convex hull, $a \in \partial A$ and $b \in \partial B$, across
the tangent to $B$ at $b$, into the $xy$-plane, surrounding the base $B$,
with the top $A$ attached to one appropriately chosen rib.
The main result of this paper is that every smooth prismatoid
has a nonoverlapping volcano unfolding (Theorem~\theoref{nonoverlap}).

If a particular unfolding of a smooth prismatoid does overlap, then
it can be converted into an overlapping unfolding of a polyhedral prismadoid, by
sufficiently fine polygonal approximations to the base and top curves of the
smooth prismatoid.  (Cf.~Fig.~\figref{mouseunfolding}.)
Thus, if there were a smooth prismatoid with no nonoverlapping
unfolding, this would imply the same result for polyhedral prismatoids.
However, the reverse implication does not hold:  Our theorem does not imply
that every polyhedral prismatoid can be unfolded without overlap.  We do hope,
however, that the smooth case will inform design of an algorithm
to handle the polyhedral case.

\section{Basic Properties}
We use $\P$ to denote a smooth prismatoid.
Its smooth convex base in the $xy$-plane is $B$, 
its smooth convex top is $A$, lying in a parallel
plane a distance $z$ above the $xy$-plane.  
$\P$ is the convex hull of $A \cup B$.
We use $A$ and $B$ to represent the curves, and, when convenient, the regions
bounded by the curves.
Thus the notation $p \in A$ should be read as $p \in \partial A$, for
we will not need to consider points interior to the region.
$A_0$ is the orthogonal projection of $A$ onto the $xy$-plane.
We place no restriction on the relationship between $A_0$ and $B$, but
it will be convenient to assume at first that $A_0 \subset B$.

We parameterize $B$ by a function $b(t)$ such that for each
$t \in [0,2\pi]$, $b(t)$ is a point on $B$.
We choose a parametrization so that $b(t)$ moves at unit velocity,
i.e., $| \dot b | = 1$.\footnote{
  The length of a vector $v$ is $|v|$.
}
Similarly, $A$ is parameterized by $a(t)$, in concert so
that the rib $(a(t),b(t)) = ab$ is a segment of $\P$.
(The dependence on $t$ often will be suppressed in contexts where it may be inferred.)
Note that the parametrization of $A$ is controlled by that of $B$
and the convex hull construction,
and would normally result in $a(t)$ moving at a variable velocity.
It is important in what follows to recognize
that $ab$ is a segment of the convex hull,
and so is the intersection of a supporting plane $H$ with $\P$,
where $H$ is tangent to both $A$ and $B$ at $a$ and $b$ respectively.
Thus the tangents of $A$ at $a$, and of $B$ at $b$, are parallel, for
these tangents lie in $H$, as well as in the planes containing $A$
and $B$ respectively.

One can define a \emph{flat prismatoid} as the shape that is the limit
of some prismatoid $\P$ of height $z$, as $z \rightarrow 0$.
For flat prismatoids, $A_0=A$.  Flat prismatoids are in a sense the
most difficult to unfold without overlap.
We will start our investigation with flat prismatoids with $A \subset B$,
and eventually remove the nesting condition, and later the flat restriction.

\paragraph{Side Unfolding.}
We define the unfolding of the side of $\P$ to be the
collection of unfolded ribs $ab$, where each rib is unfolded by rotating
it around the tangent at $b \in B$ until it lies in the $xy$-plane.
During this rotation, the angle between the rib $ab=(a(t),b(t))$ and the tangent
$\dot b (t)$ remains fixed, for this is the angle on the surface
of $\P$ at $b(t)$.
The unfolding rotation moves $a$ on the rim of the base of a right circular
cone whose apex is $b$ and whose axis is parallel to $\dot b (t)$; see Fig.~\figref{cone}.
\begin{figure}[htbp]
\centering
\includegraphics[width=0.75\linewidth]{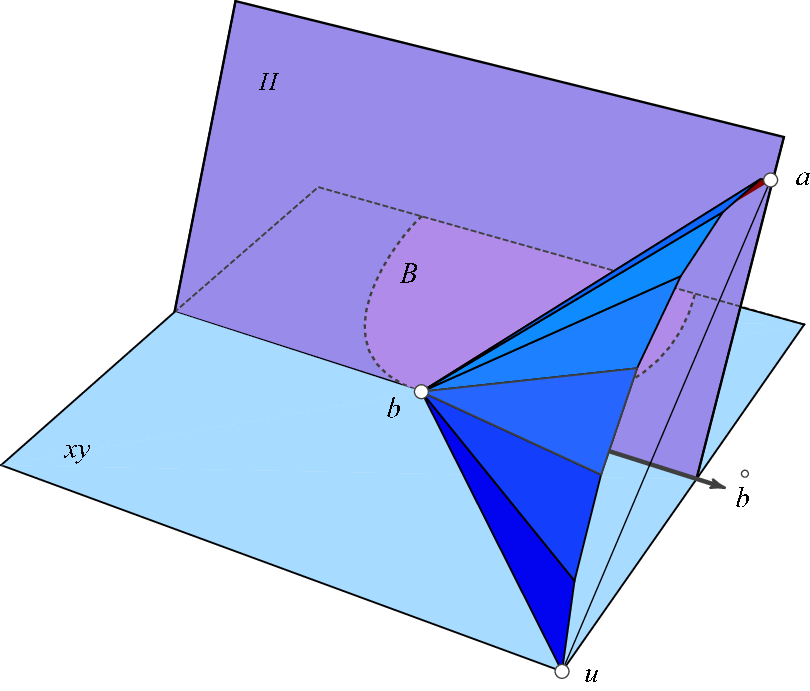}
\caption{The unfolding of rib $ab$. $H$ is the supporting plane
such that $H \cap \P = ab$.}
\figlab{cone}
\end{figure}
For flat prismatoids, $ab$ 
unfolds to a segment of the same length reflected
across $\dot b(t)$, which amounts to rotating $180^\circ$ around this cone.

We record the following observation, evident from  Fig.~\figref{cone}, for
later reference:

\begin{lemma}
For any smooth prismatoid, $u(t)-a(t)$ is always orthogonal to $\dot b(t)$.  
\lemlab{orthogbdot}
\end{lemma}
\begin{pf}
The segment $au$ lies in the circular base of the cone
in Fig.~\figref{cone}.  Any chord of the base circle is orthogonal to
the axis $\dot b$.
\end{pf}

We define the locus of the images of $a(t)$, i.e., the tips of the unfolded ribs,
as the
\emph{side unfolding} $U$ of $\P$; the qualifier ``side'' will be dropped
when clear from the context.
$U$ is parametrized as $u(t)$, such that $a(t)$ unfolds to $u(t)$.
The unfolding of a right circular cone
(Fig.~\figref{reg.pyr.stars}(c)) and a truncated right circular cone
are both circles,
but in general the unfolding can be more complicated, as
Fig.~\figref{mouseunfolding} adumbrates.
Fig.~\figref{3D.prismatoid} shows a more complex example.
\begin{figure}[htbp]
\centering
\includegraphics[width=0.9\linewidth]{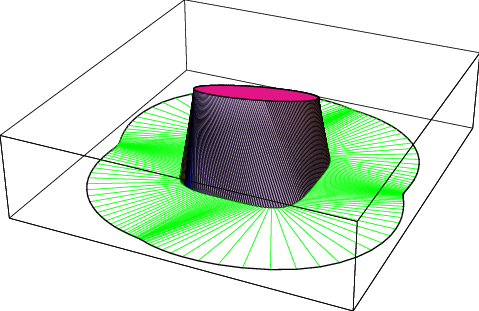}
\includegraphics[width=0.9\linewidth]{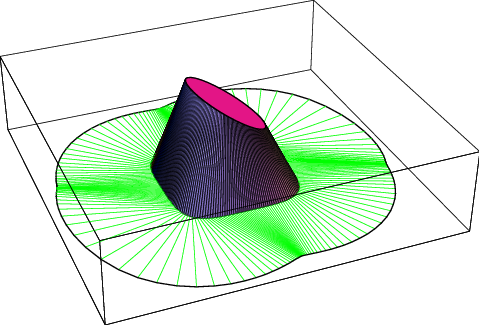}
\caption{Two views of the side unfolding of a 3D prismatoid.
The top $A$ is an ellipse in a plane
parallel to the base.
(See Fig.~\protect\figref{gallery1}a for an overhead view of a flat version of
this prismatoid.)}
\figlab{3D.prismatoid}
\end{figure}

We now argue 
that the smoothness of $A$ and $B$ implies
smoothness of $U$.
\begin{lemma}
$u(t)$ is a smooth function of $t$.
\lemlab{smooth}
\end{lemma}
\begin{pf}
We will not write out an explicit equation for $u(t)$ (except in the flat case, below),
but we can describe the form of such an equation without computing it.
The rotation shown in Fig.~\figref{cone} could be written as a matrix multiplication
that rotates $a$ through the depicted angle (call it $\theta$) about the line parallel to $\dot b$
through $b$.  This would express $u$ as a polynomial function whose terms
include $\sin \theta$, $\cos \theta$, and the components of $b$, $\dot b$, and $a$.
For smooth $A$ and $B$, all these terms are themselves smooth;
in particular $\theta(t)$ is smooth.  And because there is
no division involved in the expression, $u(t)$ is a polynomial of smooth functions,
and so is itself smooth.
\end{pf}

\noindent
In particular, this means that $u(t)$ is differentiable, which is all we need 
in the sequel.


\section{Flat Prismatoids}
Throughout this entire section, we assume $\P$ is flat, so that $A = A_0$.
We also start by assuming, in addition,
that $A \subset B$.

\subsection{Nonoverlap of the Unfolding}
We first show that $U$ itself does not self-overlap.
(We have only found a somewhat cumbersome proof of this nearly obvious fact.)
\begin{lemma}
For flat prismatoids, $U$ does not self-overlap.
\lemlab{U.nonoverlap.flat}
\end{lemma}
\begin{pf}
Suppose to the contrary that it did.  That means that
there are two ribs $a_1 b_1$ and $a_2 b_2$ whose corresponding 
unfoldings $u_1 b_1$ and $u_2 b_2$, intersect.
(Here we are using $a_1$ as an abbreviation for $a(t_1)$, etc.)
As noted above, we must have $\dot a_1$ parallel to $\dot b_1$.
Orient so that these two tangents are horizontal.  By relabeling and/or reflection if
necessary, we can arrange that $a_2$ is right of $a_1$.
By convexity of $A$, $a_2$ must be below $a_1$.

We distinguish two cases, depending on whether $b_2$ is beyond the vertical
clockwise around $B$, or not.
In terms of $t$, the cases depend on the difference $t_1 - t_2$ (where
the $t$ angles are measured counterclockwise):
\begin{enumerate}
\item Case $t_1 - t_2 \in [0,\frac{\pi}{2}]$.

See Fig.~\figref{U.nonoverlap}.
\begin{figure}[htbp]
\centering
\includegraphics[width=\linewidth]{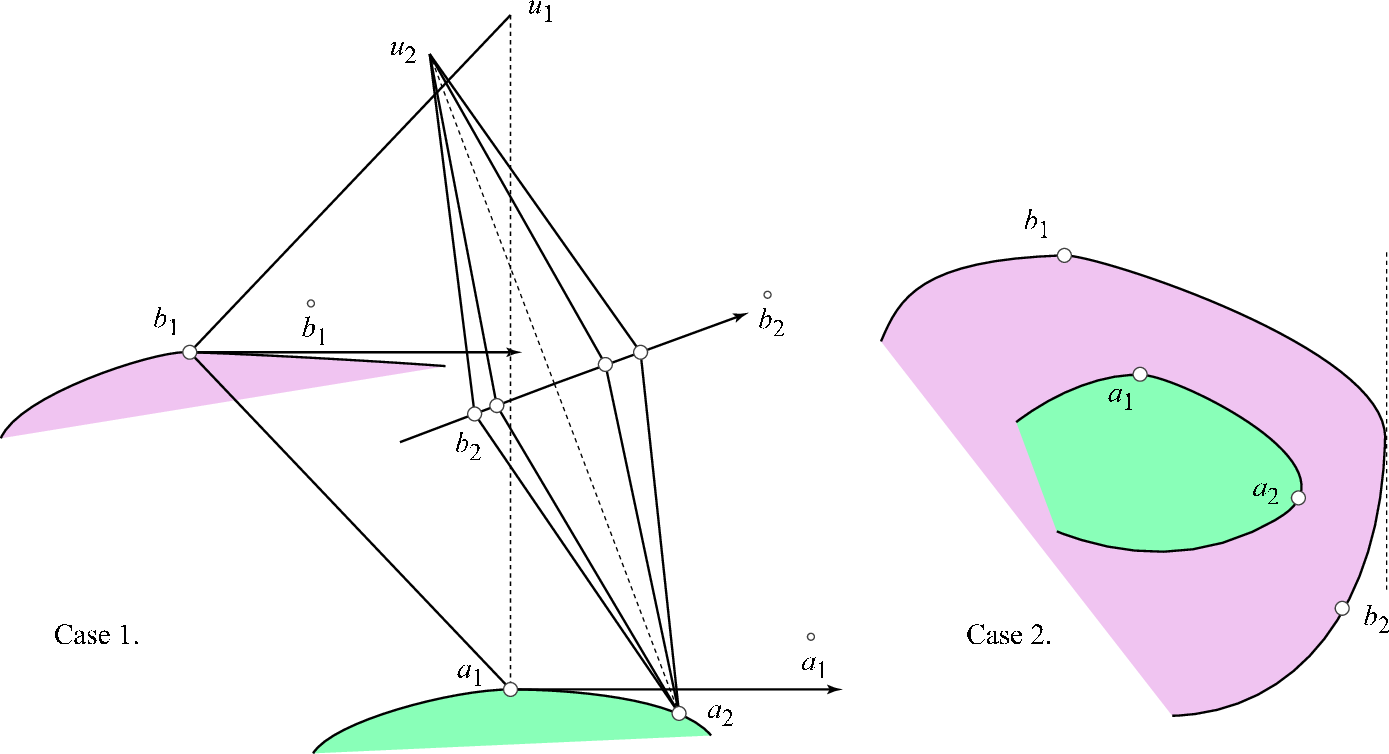}
\caption{Construction for proof of nonoverlap of $U$.}
\figlab{U.nonoverlap}
\end{figure}
Draw the line $a_2 u_2$, under the assumption that $u_1 b_1$ and $u_2 b_2$, intersect.
We must have $u_2$ left of $u_1$ in order to obtain this intersection.
But then, regardless of where $b_2$ lies, the tangent $\dot b_2$ is turned
upward (counterclockwise) with respect to $\dot b_1$, whereas convexity of $B$ demands that
it turn downward (clockwise).
Thus, intersection of these reflected ribs is incompatible with the convexity of
$A$ and $B$.
\item Case $t_1 - t_2 \in [\frac{\pi}{2},\pi]$.
$a_2$ is also placed clockwise beyond the vertical in this case.
By continuity of the tangents to the base, 
we need only look at the endpoints of the angular interval.  
If the angle between $t_1$ and $t_2$ is $\frac{\pi}{2}$, 
then $\dot b_2$ is vertical and the segment $b_2u_2$ is restricted to the halfplane 
to the right of $\dot b_2$.  
But the segment $b_1u_1$ is left of this halfplane, 
so there is no possibility for overlap.  
If the angle between $t_1$ and $t_2$ is $\pi$, 
then  $\dot b_2$ is horizontal and the segment $b_2u_2$ is restricted to the halfplane below $\dot b_2$. 
But the segment $b_1u_1$ is above this halfplane, so there is no possibility for overlap
\end{enumerate}
\end{pf}

\noindent
Note that this proof relies on convexity.
Were either $A$ or $B$ nonconvex, $U$ might well overlap.
We will see later (Lemma~\lemref{U.nonoverlap}) the lemma remains true for nonflat prismatoids.

\subsection{Tangency and Overlap}
Lemma~\lemref{U.nonoverlap.flat} shows that the only concern for volcano unfoldings
is the placement of $A$.  If one thinks of a smooth prismatoid as a limit of
approximating polyhedral prismatoids, it should be clear that $A$ should be attached
to one rib, retaining its tangent angle.  More precisely, let the tangent
to $A$ at $a$ be $\dot a$.  In the flat case, 
if $ab$ unfolds to $au$, then the attached $A$ unfolds
to a reflected image $A'$ of $A$ tangent to the reflection of $\dot a$.
This is illustrated for two ribs in Fig.~\figref{mouseunfolding}.
In the nonflat case, we imagine a rotation of $A$ about $\dot a$ until 
it lies in the supporting plane $H$ that includes $ab$, and then a rigid
rotation about the cone (cf. Fig.~\figref{cone}), which again places a reflected
copy $A'$ tangent to the rotated $\dot a$.

Because we assume $A$ is smooth, it is arbitrarily close to its tangent in a neighborhood
of any point. As we argued above, $U$ is also smooth.
Therefore, overlap between the flipout of $A$ and $U$ can only be avoided
when the reflected tangent coincides with the tangent to $U$.
Because the reflection of $\dot a$ is a reflection over $\dot b$, which is parallel to $\dot a$,
we conclude that

\begin{lemma}
A volcano unfolding avoids overlap only if $A$ is attached to a rib $ab$
that enjoys \emph{mutual tangency}:
$\dot a$ is parallel to $\dot u$.
\lemlab{mutual.tangency}
\end{lemma} 

\noindent
The reason this necessary condition is not sufficient to avoid overlap is that
it only avoids overlap locally, in a neighborhood of the attachment point $a=u$.

\subsection{Reflection Geometry.}

In the flat case, the rotation illustrated in Fig.~\figref{cone} becomes reflection.
Because we assumed the parametrization of $b$ is chosen such that
$|\dot{b}(t)|=1$ for all $t$,
the lenght of the projection of $(b-a)$ onto $\dot b$ is just $(b-a) \cdot \dot b$.
We can then find a vector from $a$ to the $\dot b$ line
as  $(b-a) + \dot b [ (b-a) \cdot \dot b]$,
as illustrated in Fig.~\figref{reflection}, and from this obtain $u$:
\begin{figure}[htbp]
\centering
\includegraphics[width=0.6\linewidth]{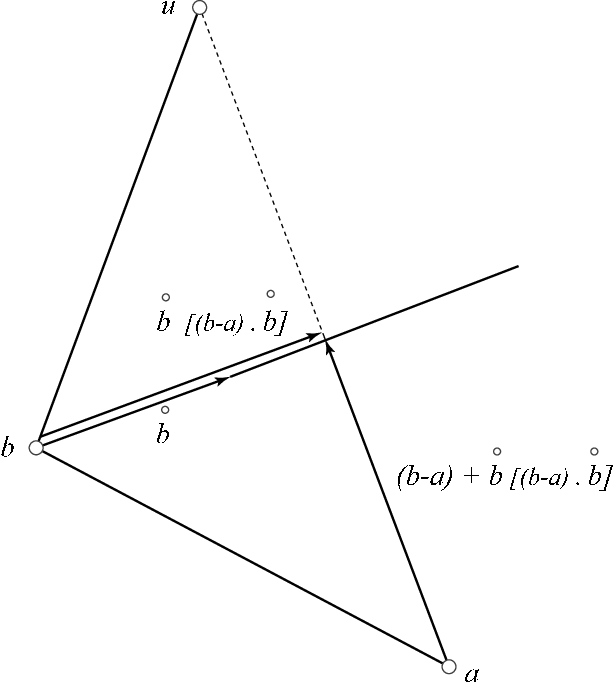}
\caption{$a$ reflects over $\dot b$ to $u$.}
\figlab{reflection}
\end{figure}
$$u(t) = a(t) + 2\{
(b(t){-}a(t)) + 
\dot{b}(t)[(b(t){-}a(t))\cdot \dot{b}(t)] \}.$$
This explicit equation for $u$ makes Lemma~\lemref{smooth}'s claim of smoothness obvious
in the flat case.

We have already seen in Fig.~\figref{mouseunfolding} that $U$ is not necessarily convex.
To give more sense of this function, we display four examples of unfoldings
of flat prismatoids
in Figs.~\figref{gallery1} and~\figref{gallery2}.
The numerical computations of the derivatives we used
to create these figures
lead to noise which produces jagged
curves and just-intersecting ribs; but with infinite precision the curves would be smooth
and the ribs nonintersecting.

\begin{figure}[htbp]
\centering
\mbox{\subfigure[Rotated ellipse inside rounded square.]{\includegraphics[width=.5\linewidth]{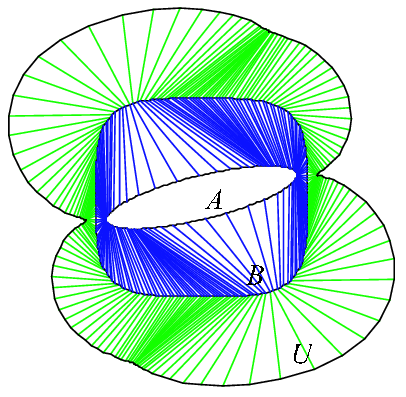}}\quad
      \subfigure[Rounded square inside mouse-shape.]{\includegraphics[width=.5\linewidth]{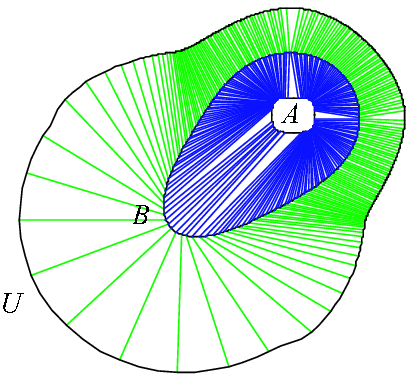}}}
\caption{Flat prismatoids.}
\figlab{gallery1}
\end{figure}

\begin{figure}[htbp]
\centering
\mbox{\subfigure[Rounded parallelogram inside rounded square.]{\includegraphics[width=.5\linewidth]{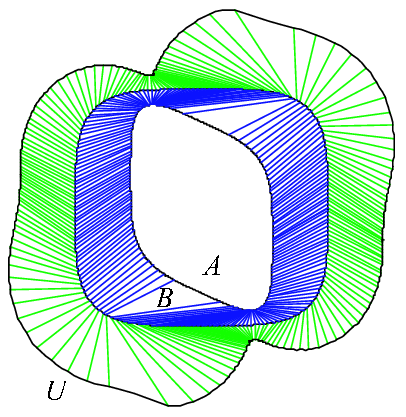}}\quad
      \subfigure[Rotated ellipse inside ellipse]{\includegraphics[width=.5\linewidth]{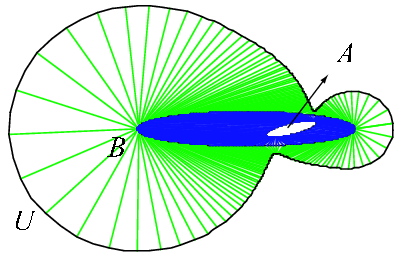}}}
\caption{More flat prismatoids.}
\figlab{gallery2}
\end{figure}

\subsection{Mutual Tangency}
\subsubsection{$A \subset B$.}
In light of Lemma~\lemref{mutual.tangency}, our goal is to find mutual
tangency between $a(t)$ and $u(t)$.  We find this tangency at the maximum distance
(in a sense)
between $A$ and $U$.

\begin{lemma}
Let $\P$ be a flat, smooth prismatoid whose top is nested inside its base.  If $|u(\hat t)-a(\hat t)|$ is at a maximum at $t=\hat t$, then mutual tangency occurs at $\hat t$.
\lemlab{mutualtangency}
\end{lemma}
\begin{pf}
Let $R(t)=|u(t)-a(t)|$ be at a maximum. Then $\dot R(t)=0$.  Note that $|u(t)-a(t)|\neq 0,$ since the top does not completely enclose the base. 

$$R(t) = \sqrt{(u(t)-a(t))\cdot{(u(t)-a(t))}}$$
$$\dot R(t)= \frac{(u(t)-a(t))\cdot{(\dot u(t)-\dot a(t))}}{\sqrt{(u(t)-a(t))\cdot{(u(t)-a(t))}}}$$
So, 
$$\dot R(t) = 0 \Longleftrightarrow (u(t)-a(t))\cdot{(\dot u(t)-\dot a(t))} = 0$$

\begin{equation}
\begin{split}
0 &= (u(t)-a(t))\cdot{(\dot u(t)-\dot a(t))}\\
  &= (u(t)-a(t))\cdot \dot u(t) + (a(t)-u(t))\cdot \dot a(t)\\
\end{split}
\end{equation} 
Thus,
$$(u(t)-a(t))\cdot \dot u(t) =(u(t)-a(t))\cdot \dot a(t).$$
Now recall that $\dot a(t)$ and $\dot b(t)$ are always parallel by
our choice of parametrization.
Second, we know from Lemma~\lemref{orthogbdot} that  $\dot b(t)$ is orthogonal
to $u(t)-a(t)$ (cf. Fig.~\figref{reflection}).
Therefore, $(u(t)-a(t))\cdot \dot a(t) = 0$.
So we have
$$(u(t)-a(t))\cdot \dot u(t) = 0$$
Because $u(t)-a(t) \neq 0$, this implies that $\dot u(t)$ is
orthogonal to $u(t)-a(t)$.  So $\dot u(t)$ and $\dot a(t)$ are 
both orthogonal to the same vector, and so parallel,
which is our definition of mutual tangency.
\end{pf}

\subsubsection{$A$ crosses $B$.}
We now remove the restriction that $A \subset B$, and permit $A$ and $B$ to ``cross.''
For flat prismatoids where $A$ and $B$ cross, some side faces face upward, and some downward.
The upward faces reflect just as before, but the downward faces do not move in the unfolding.
The next lemma shows the mutual tangency established in the previous lemma still
holds in this case.

\begin{lemma}
Let $\P$ be a flat smooth prismatoid.  
If $A$ is partially outside of $B$, 
then the global maximum of $|u(t)-a(t)|$ is on a reflected portion of the unfolding $U$. 
\lemlab{reflectedply}
\end{lemma}
\begin{figure}[htbp]
\centering
\includegraphics[width=.85\linewidth]{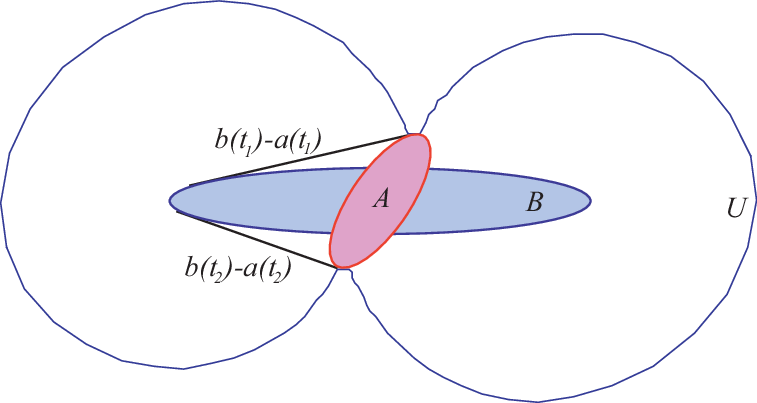}
\caption{Bi-Tangents.}
\figlab{bi-tangents}
\end{figure}
\begin{pf}
Let $A$ be partially outside of the $B$.  Then there is a bi-tangent to A and B at each transition from a reflected portion of the unfolding $U$ to a nonreflected portion. 
See Fig.\figref{bi-tangents}.
At any bi-tangent, we have $b(t)-a(t)$ coinciding with $\dot a(t)$ and $\dot b(t)$.  Thus, $a(t)$ is reflected onto itself, so $u(t)-a(t) = 0$ at any bi-tangent.  Furthermore, all nonreflected ribs have $|u(t)-a(t)|=0$. Let $U_{r}$ denote a reflected portion of the unfolding $U$.  Then $U_{r}$ is bounded by two bi-tangents, say at $t_{1}$ and $t_{2}$.  The unfolding $U$ is continuous on $[t_{1},t_{2}]$ and differentiable on $(t_{1},t_{2})$.  Thus, there is a local maximum on $(t_{1},t_{2})$, since $|u(t_1)-a(t_1)| = 0 = |u(t_2)-a(t_2)|$ and $|u(t)-a(t)|\geq 0$ for all $t \in [t_1,t_2]$.  Comparing the lengths of the local maxima obtained from the reflected portions of the unfolding $U$, there is a longest such maximum which is the global maximum of $|u(t)-a(t)|$.
\end{pf}

\begin{cor}
Mutual tangency occurs at the global maximum of $|u(t)-a(t)|$, which is always on the reflected portion of the $U$.
\corlab{nonnestedmutualtan}
\end{cor}

\subsubsection{$A \supset B$.}
Finally, we remove all restrictions on the relationship between $A$ and $B$, permitting $A$ to
enclose $B$.

\begin{lemma}
Let $\P$ be a flat prismatoid.  If $A$ completely  encloses $B$, then the tangents to $u(t)$ match the tangents to $a(t)$ and $b(t)$ for all values of $t$.
\lemlab{nestedB}
\end{lemma}
\begin{pf}
Let $A$ completely enclose $B$ and let $z=0$. Then all ribs $b(t)-a(t)$ are nonreflected.  So, $u(t)-a(t)=0$ for all $t$. Therefore, $\dot u(t)=\dot a(t)$.  This implies that the tangents to $u(t)$ match the tangents to $a(t)$ and $b(t)$ for all values of $t$, since $\dot a(t)$ and $\dot b(t)$ are always parallel.    
\end{pf}

\subsection{Offset Curve}

We have now established that the global maximum of $|u(t)-a(t)|$ yields mutual
tangency for all flat prismatoids.
But, as we observed earlier, this only means that the flip-out $A'$ does not locally
overlap $U$.  To achieve global nonoverlap, we need to prove that the
global maximum is achieved at a point on the convex hull of $U$, for then
the tangent $\dot u(\hat t)$ provides a supporting line for $U$, separating
$A'$ from $U$.
We establish this via an \emph{offset curve} (or ``parallel curve'') for $A$,
one displaced from $A$ by a constant offset along the curve's normal.
We first establish the nearly obvious claim that the offset of a convex curve is convex.

\begin{lemma}
The normals to $a(t)$ are normal to any offset curve $o(t)$ of $a(t)$.
\lemlab{offset}
\end{lemma}
\begin{pf}
Let $a(t)$ be parameterized by arc length.  
(Note, this is not the parametrization we employed before.)
Let $n(t)$ be the unit normal vectors to $a(t)$. 
Then $n(t) = (\dot{a_{2}}(t),-\dot{a_{1}}(t)).$  
Let $o(t)$ be a parallel curve of A.  
Then $o(t) = a(t)+ k n(t)$ for some constant $k$.  So,
\begin{gather}
 \begin{split}
\dot o(t)\cdot n(t) &= (\dot a(t) + k \dot{n}(t))\cdot n(t)\\
                        &= \dot a(t)\cdot n(t) + k n(t)\cdot \dot n(t)\\
                        &= k (\dot{a_{2}}(t),-\dot{a_{1}}(t))\cdot (\ddot{a_2}(t),-\ddot{a_1}(t))\\
                        &= k (\dot a(t)\cdot \ddot a(t))\\
                        &= 0\text{, since $a(t)$ is parameterized by arclength.}\\
 \end{split}
\end{gather}
Thus the normals to $a(t)$ are normal to $o(t)$. 
\end{pf}

\begin{cor}
If $a(t)$ is convex, then $o(t)$ is convex.  
\corlab{convexoffset}
\end{cor}
\begin{pf}
The normals to $a(t)$ are normal to $o(t)$. So the tangents to $a(t)$ are parallel to the tangents to $o(t)$.  Therefore, $o(t)$ is convex, since $a(t)$ is convex.  
\end{pf}

Finally, we prove the global maximum is achieved on the hull of $U$.

\begin{lemma}
If $|u(t)-a(t)|$ is a global maximum, $u(t)$ is on the convex hull of $U$.
\lemlab{prismatoidhull}
\end{lemma}
\begin{figure}[htbp]
\centering
\includegraphics[width=.65\linewidth]{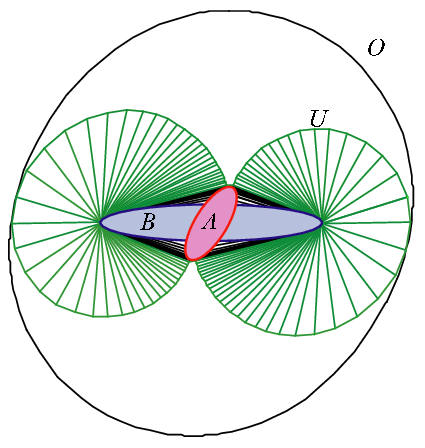}
\caption{$u(\hat t)$ is on the hull of $U$, where $O$ touches $U$.}
\figlab{crossedellipsehull}
\end{figure}
\begin{pf}
Let $|u(t)-a(t)|$ be a global maximum at $\hat t$, and let  $M=|u(\hat t)-a(\hat t)|$. Then there is mutual tangency at $\hat t$ by Lemma \lemref{mutualtangency}.  Let $n(t)$ be the unit normal vectors to $A$.  Then the offset curve, $o(t) = a(t) + M n(t)$, touches $U$ at $u(\hat t)$, because $|u(\hat t)-a(\hat t)|$ is in fact $M$, and $u(\hat t)-a(\hat t)$ is orthogonal to $\dot{a}(\hat t)$ by Lemma \lemref{offset} .  Therefore the curve $o(t)$ must enclose $U$.  For suppose some point of $U$ were outside $o(t)$.  Then its orthogonal distance from $A$ would be greater than $M$, contradicting $M$ being the maximum value of $|u(t)-a(t)|$.  By \corref{convexoffset} $o(t)$ is convex just as $A$ is.  Take a line $\ell$ tangent to $o(t)$ at $\hat t$, i.e., a line with $o(t)$ wholly to one side, since $\dot {u}(\hat t)$ is orthogonal to $u(\hat t)-a(\hat t)$.  Then $\ell$ is a supporting line to $U$.  So, $u(\hat t)$ is on the hull.
\end{pf}

\begin{cor}
There is a nonoverlapping volcano unfolding of any flat, smooth prismatoid.
\corlab{nonoverlap.flat}
\end{cor}
\begin{pf}
Flip out $A$ attached to the rib $(a(\hat t), b(\hat t))$ for the $\hat t$ that
achieves the global maximum of $|u(t)-a(t)|$.
Lemma~\lemref{mutual.tangency} guarantees mutual tangency,
and Lemma~\lemref{prismatoidhull} guarantees this tangency occurs on the hull of $U$.
\end{pf}

\section{Nonflat Smooth Prismatoids}

We have concentrated on flat prismatoids for two reasons:
the geometric relationships are clearer, and in some sense, flat prismatoids are the most
difficult.  Roughly speaking, lifting $A$ of a flat prismatoid to $z > 0$ rounds
out the unfolding $U$, and maintains all the key relationships we need.
If we imagine a continuous lifting from $z=0$, then initially $u=u_0$,
and then $u$ moves out along the line through $a_0$ and $u_0$.
See Fig.~\figref{nonzeroZ}.
This is because the lifting can be seen as a widening of the cone shown 
in Fig~\figref{cone}, while maintaining the cone base cutting the $xy$-plane
along the same line.
These relationships permit the extension of Lemma~\lemref{U.nonoverlap.flat}:

\begin{lemma}
For nonflat prismatoids, $U$ does not self-overlap.
\lemlab{U.nonoverlap}
\end{lemma}
\begin{pf}
The collinearity illustrated in Fig.~\figref{nonzeroZ}
implies that the geometric situation illustrated in
Fig.~\figref{U.nonoverlap} for the flat case still holds with only minor variation.
In particular, we still have triangles $\triangle a_1 b_1 u_1$ etc. as in
that figure, but now the triangles are extended out to the true $u_1$ 
(i.e., $u(t_1)$ rather than the projected/flat $u_0(t_1)$; cf.~Fig.~\figref{nonzeroZ} ), etc.
Because our reasoning in Lemma~\lemref{U.nonoverlap.flat}
only depended on the triangles, and not that they were isosceles,
all else remains the same, and the
assumption of overlap again contradicts convexity of $A$ and $B$.
\end{pf}

\subsection{Distance Maximum and Mutual Tangency}
The central burden of this section
is to show that the global maximum of $|u-a|$ for the nonflat case
is achieved at the same $\hat t$ as in the flat case.
We prove this in two steps: from $|u_0-a_0|$ to $|u-a_0|$ (Lemma~\lemref{Part1 projectedmax}),
and from  $|u-a_0|$ to $|u-a|$ (Lemma~\lemref{Part2 projectedmax}).

\begin{lemma}
Let $\P$ be a nonflat smooth prismatoid with the projection of $A$ nested in $B$.  
Let $u_{0}(\hat {t})$ denote the unfolding obtained from the projection of $a(\hat {t})$ 
onto the $xy$-plane of the base at time $\hat {t}$.  
This projection will be denoted by $a_{0}(\hat {t})$.  
If $|u_{0}(\hat {t})-a_{0}(\hat {t})|$ is a global maximum, 
then $|u(\hat {t})-a_{0}(\hat {t})|$ is a global maximum.
\lemlab{Part1 projectedmax}   
\end{lemma}
\begin{figure}[htbp]
\centering
\includegraphics[width=.75\linewidth]{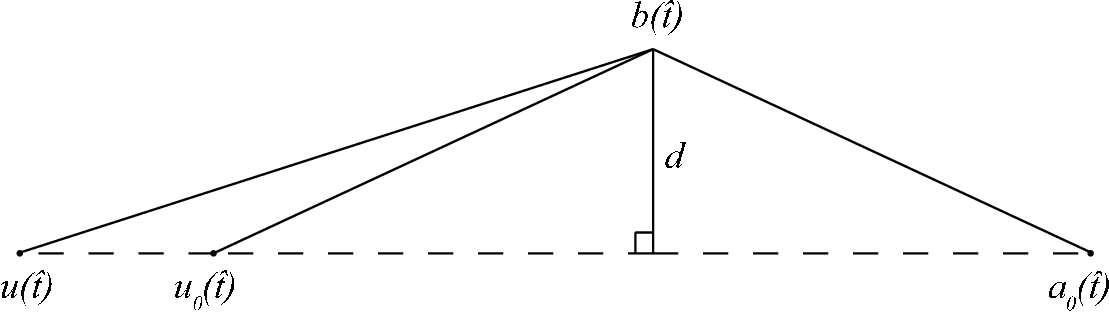}
\caption{Lifting $A$ to $z>0$ maintains collinearity of
$u$, $u_0$, and $a_0$.}
\figlab{nonzeroZ}
\end{figure}
\begin{pf}
Let $|u_{0}(\hat {t})-a_{0}(\hat {t})|$ be a global maximum at $\hat t$.  
Let $\ell$ denote the line through $u(\hat {t})$, $u_{0}(\hat {t})$, and $a_{0}(\hat {t})$ and let $d$ denote the altitude displayed in Fig.~\figref{nonzeroZ}.  
Then $d=\sqrt{|b(\hat {t})-a_{0}(\hat {t})|^2-(\frac{1}{2}|u_{0}(\hat {t})-a_{0}(\hat {t})|)^2}$.\\  
By Figure \figref{nonzeroZ}, we see that $|u(\hat {t})-u_{0}(\hat {t})|=|proj_{\ell}(u(\hat {t})-b(\hat {t}))|-|proj_{\ell}(u_{0}(\hat {t})-b(\hat{t}))|.$\\
We can further expand this expression by using the following two equalities:
\begin{align}|proj_{\ell}(u(\hat {t})-b(\hat {t}))|&=\sqrt{|u(\hat {t})-b(\hat {t})|^2-d^2}\\
\mbox{} &= \sqrt{|b(\hat {t})-a(\hat{t})|^2-d^2}=\sqrt{|b(\hat {t})-a_{0}(\hat {t})|^2+z^2-d^2}\\
|proj_{\ell}(u_{0}(\hat {t})-b(\hat{t}))|&=\sqrt{|u_{0}(\hat {t})-b(\hat{t})|^2-d^2}=\sqrt{|b(\hat {t})-a_{0}(\hat {t})|^2-d^2}.
\end{align}
So,
\begin{gather}
\begin{split}
|u(\hat {t})-u_{0}(\hat {t})|&=\sqrt{|b(\hat {t})-a_{0}(\hat {t})|^2+z^2-d^2}-\sqrt{|b(\hat {t})-a_{0}(\hat {t})|^2-d^2}\\
&=\sqrt{|b(\hat {t})-a_{0}(\hat{t})|^2+z^2-(|b(\hat {t})-a_{0}(\hat {t})|^2-(\frac{1}{2}|u_{0}(\hat {t})-a_{0}(\hat {t})|)^2)}\\&-\sqrt{|b(\hat {t})-a_{0}(\hat{t})|^2-(|b(\hat {t})-a_{0}(\hat {t})|^2-(\frac{1}{2}|u_{0}(\hat {t})-a_{0}(\hat {t})|)^2)}\\
&=\sqrt{z^2+\frac{1}{4}|u_{0}(\hat {t})-a_{0}(\hat {t})|^2}-\sqrt{\frac{1}{4}|u_{0}(\hat {t})-a_{0}(\hat {t})|^2}
\end{split}
\end{gather}

Thus,
\begin{gather}
\begin{split}
|u(\hat{t})-a_{0}(\hat{t})| &= |u(\hat {t})-u_{0}(\hat {t})| + |u_{0}(\hat{t})-a_{0}(\hat{t})|\\
                            &= \sqrt{z^2+\frac{1}{4}|u_{0}(\hat {t})-a_{0}(\hat {t})|^2} + \frac{1}{2}|u_{0}(\hat {t})-a_{0}(\hat {t})|
\end{split}
\end{gather}
Because $z$ is fixed independent of $t$,
$|u(\hat{t})-a_{0}(\hat{t})|$
is a global maximum.
\end{pf}

\begin{lemma}
If $|u(\hat{t})-a_{0}(\hat{t})|$ is a global maximum, then $|u(\hat{t})-a(\hat{t})|$ is a global maximum.  
\lemlab{Part2 projectedmax}
\begin{figure}[htbp]
\centering
\includegraphics[width=.5\linewidth]{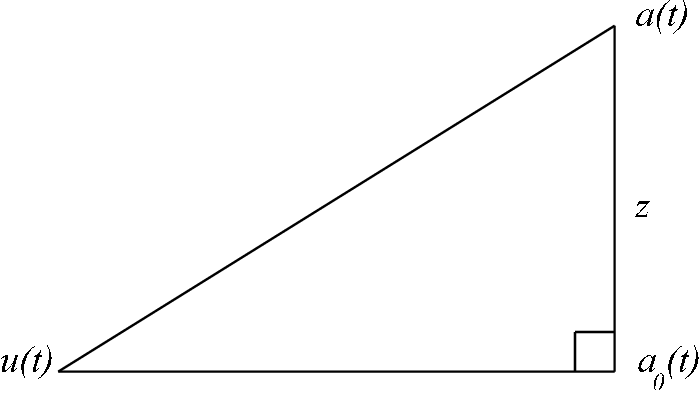}
\caption{$A$ lifted by $z$.}
\figlab{nonzeroPAmax}
\end{figure}
\end{lemma}
\begin{pf}
Let $|u(\hat{t})-a_{0}(\hat{t})|$ be a global maximum at $\hat{t}$.  Recall that 
(see Fig.~\figref{nonzeroPAmax})
$$|u(\hat{t})-a(\hat{t})| = \sqrt{|u(\hat{t})-a_{0}(\hat{t})|^2 + z^2},$$ 
which is a global maximum since $z$ is fixed for all $t$.   
\end{pf}


\begin{lemma}
Consider a nonflat smooth prismatoid of height $z>0$.  
If $A$ is partially outside of $B$, 
then there is a global maximum of $|u(t)-a_{0}(t)|$ on the reflected portion of the unfolding $U$.  
\lemlab{z>0,non-nested}
\end{lemma}
\begin{pf}
Let $A$ be partially outside of $B$.  Then the projection $A_{0}$ is partially outside of $B$.  By Lemma \lemref{reflectedply}, the global maximum of $|u_{0}(t)-a_{0}(t)|$ is on a reflected portion of $U_{0}$ say at $\hat t$.  So, $|u(\hat t) - a_{0}(\hat t)|$ is a global maximum.  ( Note that this implies that $|u(\hat t) - a(\hat t)|$ is a global maximum by Lemmas \lemref{Part1 projectedmax} and \lemref{Part2 projectedmax}.)  
\end{pf}

\begin{lemma}
If $A$ completely encloses $B$ and $z>0$, then mutual tangency occurs at the global maximum of $|u(t)-a_{0}(t)|$.
\lemlab{mutualtanz>0}
\end{lemma}
\begin{pf}
Let $A$ completely enclose $B$ and let $R(t) = |u(t)-a_{0}(t)|$ obtain its global maximum at $\hat t$.  Then $\dot R(\hat t)=0$.  Note that $|u(t)-a_{0}(t)|\neq 0,$ since $A_{0}$ and $B$ are noncrossing.  
$$R(t) = \sqrt{(u(t)-a_{0}(t))\cdot{(u(t)-a_{0}(t))}}$$
$$\dot R(t)= \frac{(u(t)-a_{0}(t))\cdot{(\dot u(t)-\dot a_{0}(t))}}{\sqrt{(u(t)-a_{0}(t))\cdot{(u(t)-a_{0}(t))}}}$$
So, 
$$\dot R(\hat t) = 0 \Longleftrightarrow (u(\hat t)-a_{0}(\hat t))\cdot{(\dot u(\hat t)-\dot a_{0}(\hat t))} = 0$$

\begin{equation}
\begin{split}
0 &= (u(\hat t)-a_{0}(\hat t))\cdot{(\dot u(\hat t)-\dot a_{0}(\hat t))}\\
  &= (u(\hat t)-a_{0}(\hat t))\cdot \dot u(\hat t) + (a_{0}(\hat t)-u(\hat t))\cdot \dot a_{0}(\hat t)\\
\end{split}
\end{equation} 
Thus,
$$(u(\hat t)-a_{0}(\hat t))\cdot \dot u(\hat t) =(u(\hat t)-a_{0}(\hat t))\cdot \dot a_{0}(\hat t).$$
But since $u(\hat t)-a_{0}(\hat t) \neq 0$,
$$\dot u(\hat t)= \dot a_{0}(\hat t).$$ 
But $\dot a_{0}(\hat t)$ is parallel to $\dot a(\hat t)$, so mutual tangency must occur here at $\hat t$.  
\end{pf}

Fig.~\figref{nonflatoffset} illustrates that $\hat t$ achieving the maximum
for the flat case corresponds
to maximum in the nonflat case.

\subsection{Offset Curve}
We again use an offset of $A$ to prove that the above identified maximum occurs on the convex hull
of $U$.
\begin{lemma}
If $|u(t)-a_{0}(t)|$ is a global maximum at $\hat t$, then $u(\hat t)$ is on the convex hull of $U$.  
\end{lemma}
\begin{figure}[htbp]
\centering
\includegraphics[width=.75\linewidth]{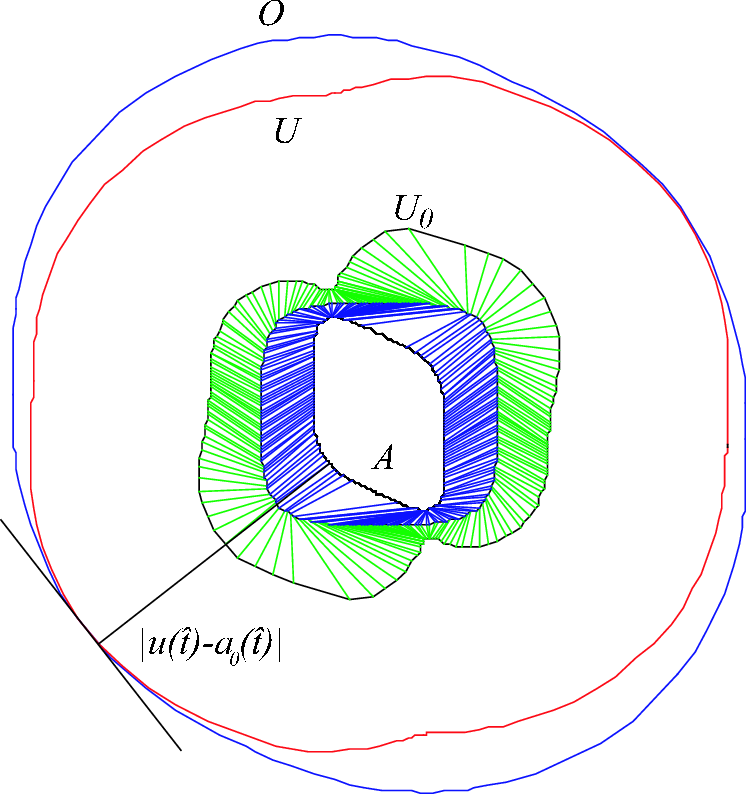}
\caption{$U_0$ is the unfolding of the flat prismatoid, $U$ the unfolding of a lifting
of $A$ to form a nonflat prismatoid.  $O$ is the offset of $A$.}
\figlab{nonflatoffset}
\end{figure}
\begin{pf}
Let $o(t) = a_{0}(t) + |u(\hat t) - a_{0}(\hat t)| n(t)$.  
Because $u(\hat t),u_{0}(\hat t),a_{0}(\hat t)$ are collinear,
and since $u_{0}(t)-a_{0}(t)$ is orthogonal to $\dot{a}_{0}(t)$, 
we have $u(\hat t)$ orthogonal to $\dot{a}_{0}(\hat t)$.  
As in Lemma \lemref{prismatoidhull}, $o(t)$ touches $U$ at $u(\hat t)$ and  must enclose $u(t)$.   
We again take a line $\ell$ tangent to $o(t)$ at $\hat t$, i.e., a line with $o(t)$ wholly to one side.  Then $\ell$ is a supporting line to $U$.  So, $u(\hat t)$ is on the hull of $U$. 
\end{pf}

\begin{theorem}
There is a nonoverlapping volcano unfolding of any smooth prismatoid.
\theolab{nonoverlap}
\end{theorem}

\noindent
The proof is identical to that of Corollary~\corref{nonoverlap.flat},
relying on the corresponding nonflat lemmas.

\section{Discussion}
We have established that, for any smooth prismatoid $\P$,
there is always at least one spot to flip out the top $A$
so that it does not overlap with $U$, thus producing a nonoverlapping volcano unfolding.
We know of examples where there are only two such ``safe'' flip-out spots,
symmetrically placed equal global maxima of $|u-a|$.
We hope to use our analysis of smooth prismatoids to answer the question of whether or not
every polyhedral prismatoid has a nonoverlapping volcano edge-unfolding.

\paragraph{Acknowledgements.}
We benefitted from discussions with 
Erik Demaine,
Meghan Irving,
Molly Miller,
and
Gail Parsloe.

\bibliographystyle{alpha}
\bibliography{Prismatoids}
\end{document}

%% file: mac.tex
\newenvironment{pf}{\unskip{\bf Proof:}}{\unskip{\hfill $\Box$}}

\newcommand{\lemlab}[1]{\label{lemma:#1}}
\newcommand{\theolab}[1]{\label{theo:#1}}

\newcommand{\corlab}[1]{\label{cor:#1}}

\newcommand{\figlab}[1]{\label{fig:#1}}
\newcommand{\seclab}[1]{\label{section:#1}}

\newcommand{\lemref}[1]{\ref{lemma:#1}}

\newcommand{\theoref}[1]{\ref{theo:#1}}

\newcommand{\corref}[1]{\ref{cor:#1}}

\newcommand{\figref}[1]{\ref{fig:#1}}

\newtheorem{theorem}{Theorem}[section]
\newtheorem{lemma}[theorem]{Lemma}
\newtheorem{cor}[theorem]{Corollary}

\newtheorem{df}{Definition}[section]

{\catcode`\@=11
\gdef\setft#1#2#3{%
\def\@oddfoot{
{\setbox0=\hbox{#1}
\setbox1=\hbox{#3}
\ifdim\wd0>\wd1
\dimen0=\wd0
\box0\hfil#2\hfil\hbox to\dimen0{\hfil\hfil\box1}
\else \dimen0=\wd1
\hbox to\dimen0{\box0\hfil }\hfil#2\hfil\box1 \fi
}}} }

\def\complaint#1{}
\def\withcomplaints{
\newcounter{mycomplaints}
\def\complaint##1{\refstepcounter{mycomplaints}%
\ifhmode%
\unskip%
{\dimen1=\baselineskip \divide\dimen1 by 2 %
\raise\dimen1\llap{\tiny -\themycomplaints-}}\fi%
\marginpar{\tiny [\themycomplaints]: ##1}}%
}

%% file: Prismatoids.bbl
\begin{thebibliography}{O'R00}

\bibitem[DO04]{do-fucg-04}
Erik~D. Demaine and Joseph O'Rourke.
\newblock {\em Folding and Unfolding in Computational Geometry}.
\newblock 2004.
\newblock Monograph in preparation.

\bibitem[O'R00]{o-fucg-00}
Joseph O'Rourke.
\newblock Folding and unfolding in computational geometry.
\newblock In {\em Discrete Comput. Geom.}, volume 1763 of {\em Lecture Notes
  Comput. Sci.}, pages 258--266. Springer-Verlag, 2000.
\newblock Papers from the {\em Japan Conf. Discrete Comput. Geom.}, Tokyo, Dec. 1998.

\end{thebibliography}
